# Hesiod's calendar and the star Spica[1]

Elio Antonello
INAF – Osservatorio Astronomico di Brera
SIA – Società Italiana di Archeoastronomia
elio.antonello@brera.inaf.it

**Abstract.** In Hesiod's calendar, circa 8$^{th}$ century BCE, the harvest times of cereals were indicated by the heliacal rising of Pleiades (harvest) and by that of Orion (thresh). We tried to verify which risings and settings of the brightest stars could have been used as indicators in the previous millennia, taking into account the precession and the dependence of the heliacal dates on the latitude. In the second half of the 9$^{th}$ millennium BCE there was essentially one bright star that could be used both for the harvest (heliacal setting of the star) and for the thresh (heliacal rising of the star): Spica, i.e. *ear* (in Latin) of cereals. According to archaeologists, the domestication of barley and wheat occurred in Near East at the end of the 9$^{th}$ millennium BCE. Given the importance of the bright stars and asterisms for ancient farming activities, we have therefore proposed that the identification of the star α Virginis with an ear should date back to the beginning of Neolithic, possibly well before the identification of the constellation (Virgo) with a maiden.

## Introduction

The change from the life of hunters and gatherers of Palaeolithic to that based on farming activity of Neolithic represents an extraordinary revolution in the human history. It is actually in the Neolithic that we find the roots of the present state of the human race, not only in its domination and exploitation of the environment, but also in the very foundations of our culture and mentality (CAUVIN, 2000). During the Neolithic our ancestors understood the close relation between the times of the celestial phenomena (calendar) and the effectiveness of the agricultural techniques. The risings and settings of stars, along with the Sun and Moon, were used to determine the dates of the farming activities. An historic example is given by the ancient poem Ἔργα καὶ ἡμέραι *(Works and Days)* of Hesiod (Bœotia, 8$^{th}$ century BCE), in which stars and asterisms such as Arcturus, Pleiades and Orion were adopted to indicate the dates of the various activities along the year.

The name of the star α Virginis = Spica, i.e. *ear* in Latin, is suggestive of a possible relation with the farming activities related to harvest in spring/summer; however, if we consider the presumable farmer calendars of the past few millennia and the dates of the heliacal rising and setting of Spica, such a name does not seem to be justified. In fact, at mean latitudes, the dates are in the period between november (heliacal rising) and september (ten months later; heliacal setting), and the star is not visible during about fifty days between september and november. Here we will try to discuss this case, and we will propose a possible solution using as a reference the poem of Hesiod.

## Risings and settings

There are some possible misunderstandings related to the different meaning of heliacal, acronychal (or achronic, achronal, achronical) and cosmical (or cosmic) risings and settings of stars adopted by different scholars, as discussed by KELLEY & MILONE (2005). We will adopt the following definitions. Heliacal rising of a star is when it rises just before the sunrise, and heliacal setting when it sets just after the sunset. Acronychal rising of the star is when it rises just after the sunset, and acronychal setting when it sets just before the sunrise.

---

[1] Based on talks given at SEAC 2009 meeting in Alexandria (Egypt), and at archaeoastronomy meetings and conferences in Italy in 2009 - 2012.

SCHAEFER (1985) proposed a computer program for estimating the date of the heliacal phenomena. This was an objective and more scientific approach with respect to previous works. However, the model is not free of some arbitrariness in the choice of the parameters such as visibility conditions, magnitude limits and sky transparency, when it is applied to the sky of some millennia ago in some remote place, since the true conditions are not known. Therefore the estimated date must be considered in any case as an approximated one. We have rewritten the program in order to include also the acronychal phenomena (for the problems of the acronychal risings, see SCHAEFER, 1987), since they were considered in Hesiod's calendar. We adopted the photometric conditions, stellar visibility conditions and solar longitudes (present solar year) defined by Schaefer. A better sky brightness model based on the data of KOOMEN ET AL. (1952) was introduced by using the methods of the statistical regression analysis. The coordinates of the star corrected for the precession were given as input, while the number of days elapsed from the spring equinox was the output. We plan to improve the program in the future by adopting a longitude of the Sun depending on the epoch, and a better model for the precession.

**The calendar**

Analysis of Hesiod's calendar were performed by several authors, such as SCHIAPARELLI (1892) and AVENI (1989). However, Aveni did not discriminate between heliacal and acronychal phenomena, so a reader may have some difficulty in understanding his results. Here we discuss the results of our program when applied to that calendar. Hesiod declares: Πληιάδων Ἀτλαγενέων ἐπιτελλομενάων ἄρχεσθ' ἀμέτου (*When the Pleiades, daughters of Atlas, are rising, begin the harvest*). At the mid 8$^{th}$ century BCE in Bœotia, the heliacal rising of the Pleiades (assuming a magnitude 2.0 for the cluster; e.g. SCHAEFER, 1987) occurred about 47 days after spring equinox, which correspond approximately to the first week of May in the present day calendar. We adopted a mean latitude of 38 degrees for Bœotia (Figure 1). Hesiod did not specify the nature of the cereal, whether barley or wheat; this has some importance since barley ripens before wheat. However, one should note that harvest usually begun when the cereal (particularly barley) was still green, in order to avoid the loss of grains, and the ears were collected in sheaves and carefully guarded till complete drying. In Mesopotamia, owing to the poor nature of soils, mainly barley was cultivated; in other regions both barley and wheat were grown, and this could be the case also of Bœotia.

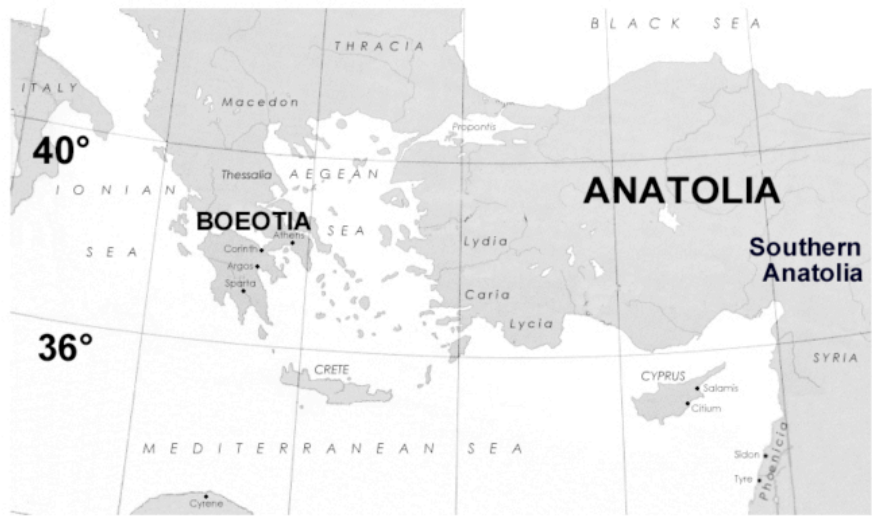

**Fig. 1.** The latitude of the regions of Bœotia and of Southern Anatolia.

According to Hesiod, threshing and winnowing in Bœotia begun when Orion risen: δμωσὶ δ᾽ ἐποτρύνειν Δημήτερος ἱερὸν ἀκτήν δινέμεν, εὖτ᾽ ἂν πρῶτα φανῇ σθένος Ὠρίωνος (*Set your slaves to winnow Demeter's holy grain, when strong Orion first appears*). The heliacal rising of the first bright stars of Orion, such as Bellatrix, occurred about 85 days after spring equinox, which correspond approximately to mid June in the present day calendar. That is, the thresh was more than a month (about 40 days) after the harvest. This could be a bit surprising since today we are familiar with farming machines that both harvest and thresh simultaneously, i.e. they cut the dried wheat and separate the grains.

An interesting confirmation of the time separation between harvesting and threshing in the past millennia can be drawn from the biblical book of Ruth and the Hebraic traditions. According to the Bible, Naomi and Ruth arrived in Bethlehem from Moab at the time of barley harvesting. That was around Pesach, usually in April (present day calendar), that is, quite early; the different date from the one for the harvest in Bœotia could be explained by the different latitude. Ruth gleaned after the harvesters with the maiden of Boaz for several days until the end of the harvest of both barley *and* wheat. After some time, Naomi talked to Ruth about the foreseen threshing and winnowing in the threshing floor of Boaz. This was also the occasion for a feast, and all the activity regarding barley and wheat should have been finished before Shavuot, that is fifty days after Pesach.

**Spica**

Pleiades and Orion could have been probably used as indicators for harvesting and threshing in Bœotia for several centuries during the 1$^{st}$ millennium BCE. Owing to the precession, however, other stars should have been adopted before and after that epoch. The question now is whether and when Spica (Figure 2) could have been one of such stars. We got a very interesting indication from the program. During the second half of the 9$^{th}$ Millennium BCE it would have been possible to use just α Virginis for both harvesting and threshing, since the date of its heliacal setting was about 47 days after spring equinox, and the date of its heliacal rising was about 85 days after spring equinox. Probably this is not a significant result for the prehistory of Bœotia, since the 9$^{th}$ Millennium BCE appears a too early epoch, however it could be of some importance for other regions in the Near East.

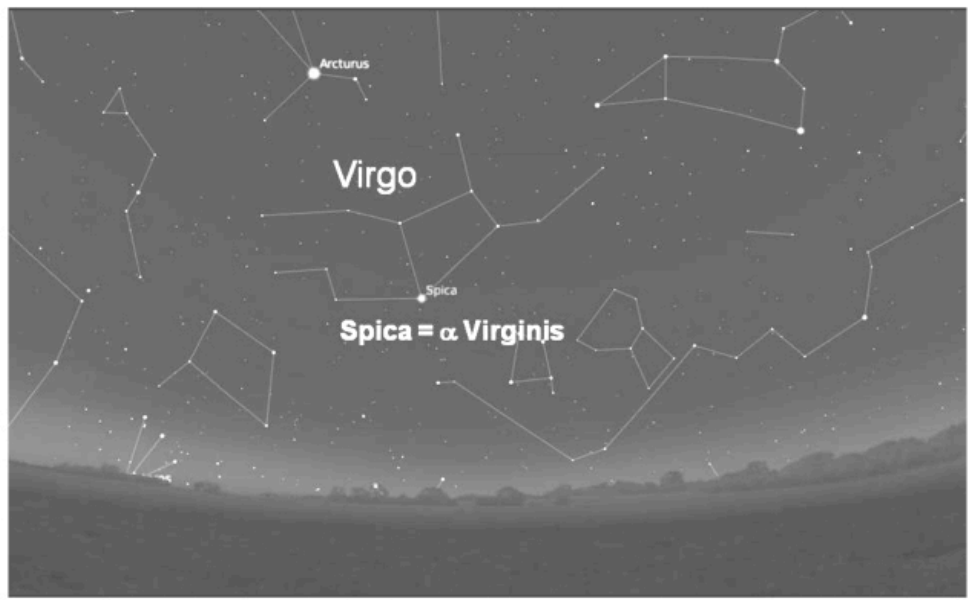

**Fig. 2.** The Virgo constellation and the star Spica (simulation with the program STELLARIUM).

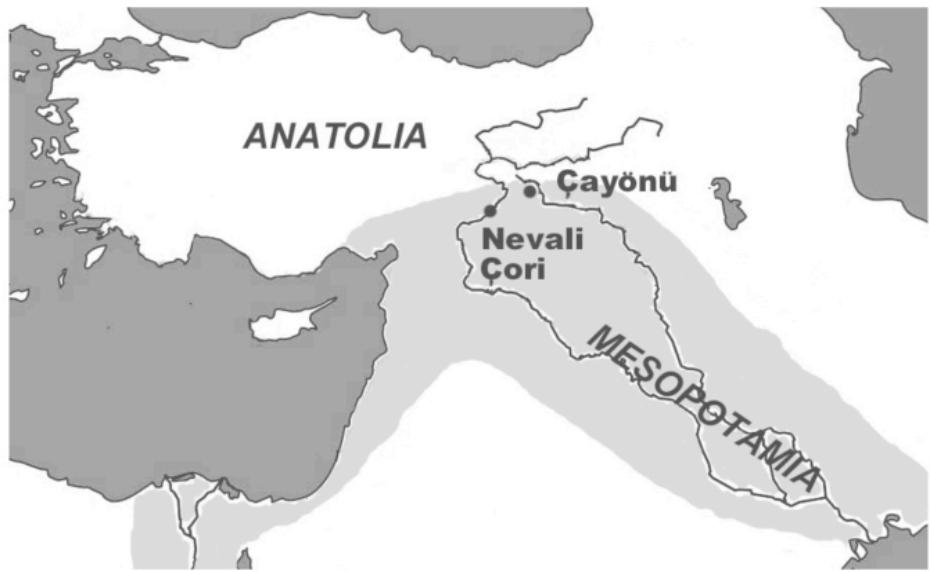

**Fig. 3.** The light grey area in the map represents the Fertile Crescent. The latitude of the Neolithic sites of Nevali Çori and Çayönü is about 37 and 38 degrees, respectively.

The latitude of central Bœotia is about 38 degrees, which is also the latitude of Southern Anatolia (Figures 1 and 3). It is not possible to know exactly the ancient climate in the two regions, and today the climate in Southern Anatolia is presumably more arid than several millennia ago, when this region was part of the Fertile Crescent. We presume a similarity of climate between the two regions in past millennia. According to this hypothesis, the times of the harvest and thresh should have been approximately the same both in Bœotia and Southern Anatolia. On the second half of the 9$^{th}$ Millennium BCE (calibrated radiocarbon dates) there was not yet a developed agriculture, though the process of the domestication of cereals was taking place in the Fertile Crescent (see e.g. CAUVIN, 2000). Seeds of domesticated cereals have been found for example in Pre-Pottery Neolithic A and Early B villages of Southern Anatolia (Figure 3) such as Çayönü and Nevali Çori (NESBITT, 2002; about 8$^{th}$ Millennium BCE, uncalibrated dates), and wild cereals were cultivated quite earlier at lower latitudes in the Jordan valley. We could think therefore that people in the Near East found the best dates for harvesting and threshing at the beginning of Neolithic, and associated such dates with the star α Virginis, that was consequently identified with an ear, Spica.

It is interesting to remark the difference of some weeks for the harvest between the Hesiod's calendar and the Hebraic traditions (that were based however on a lunar calendar), that could be explained by the different latitude, i.e. different insolation conditions and climate. We can note a sort of a correlation between the date of the heliacal setting of Spica, the latitude and the epoch. An earlier date could have been valid for harvesting at a lower latitude in an earlier epoch, such as mid-April in Palestine on the 11$^{th}$ Millennium, and a later date for harvesting at a higher latitude in a later epoch, such as beginning of May in Southern Anatolia on the 9$^{th}$ Millennium. The progressive change of the date with the epoch is due to the astronomical precession. The rising (and not the setting) of Spica could have been used as an indicator for harvesting at a very late epoch, about the 5$^{th}$ Millennium BCE, in central Europe (latitude of about 50 degrees), right at the beginning of the Neolithic civilization in that European region (ANTONELLO, 2009). This is a reasonable estimate if we take into account the present day farming activity dates at such northern latitudes. Given a possible importance of Spica for the agriculture, one could wonder whether the effect of the precession on this star may have also played some role in the diffusion of the Neolithic civilization (for the problem of diffusion in Neolithic, see e.g. CAUVIN, 2000).

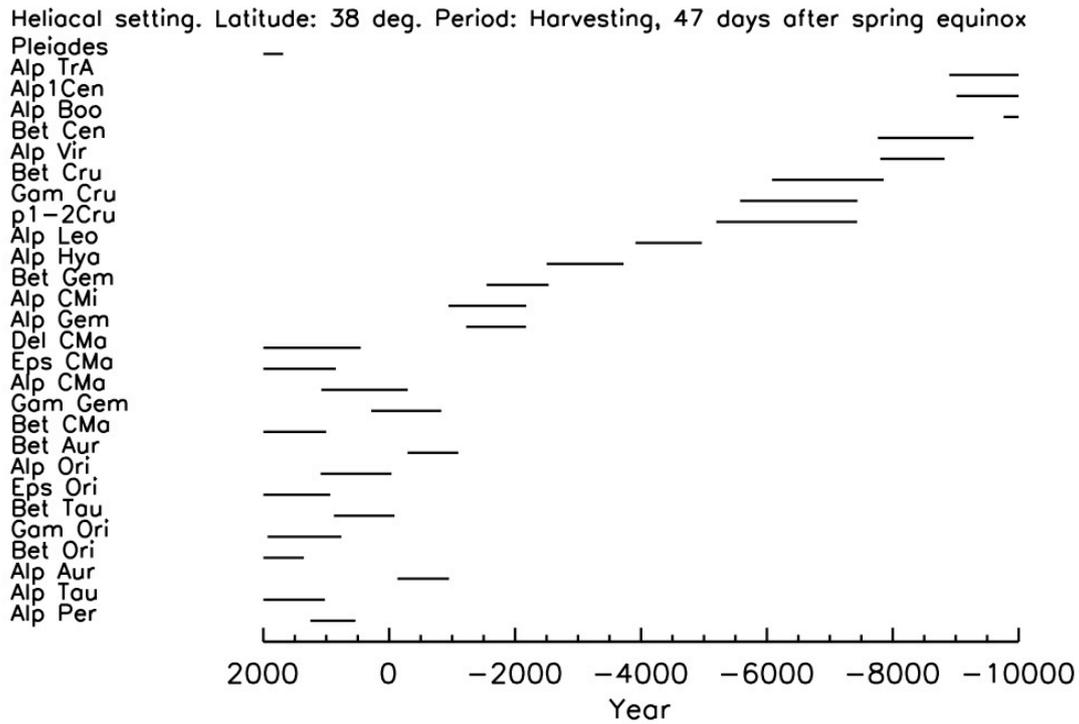

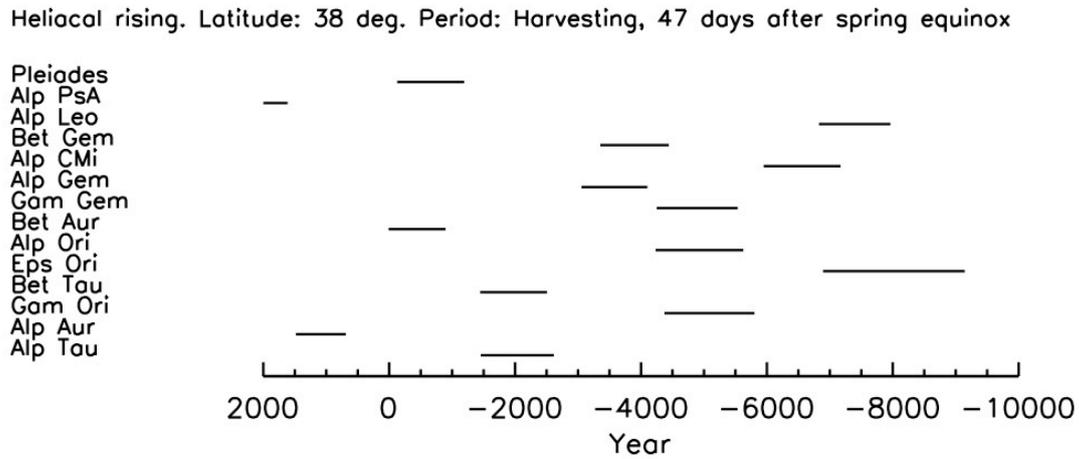

**Fig. 4.** The possible period of visibility of the heliacal setting and heliacal rising of the bright stars in the past millennia, for a latitude of 38 deg, during a harvesting time between 40 and 54 days after spring equinox (see text).

**Other stars**

As a check we have verified which heliacal (or acronychal) phenomena of stars brighter than the magnitude 2.0 could be used instead those of α Virginis during the 9[th] Millennium BCE at a latitude of about 38 degrees. We considered a period of one week before and after the given dates, that is for example from 40 to 54 days after the spring equinox as a useful time in the case of harvesting (Fig. 4). This time interval includes the date of the heliacal rising of just one star, ε Ori, but it should be practically excluded since the declination is less than –40 degrees, that is its zenith distance is larger than 80 degrees, and it was visible just for a very short time. The time interval includes also the heliacal setting of β Centauri besides that of α Virginis, and the star has a declination of –19 degrees. There are no useful acronychal risings of bright stars, while there is the acronychal setting of just one star, α Aquilæ (Altair). The conclusion of this test is that one could choose among only very few bright stars for the indication of the harvesting date during the 9$^{th}$ Millennium BCE. For the threshing date the conclusion is in part similar.

**Conclusion**

We think that during the 9$^{th}$ Millennium the people in the Fertile Crescent could have used the star α Virginis as an indicator of the time both of the harvest of cereals (in spring) and of the thresh (about forty days after harvesting). This star could have been of some help for the domestication of wild cereals and the development of farming techniques, and we think this could be the origin of its name, Spica, that is ear. The Babylonians identified α Virginis as the ear of the goddess Shala (ear of barley), while part of the present day Virgo (= maiden; Figure 5) constellation was identified by them not with a maiden but with the furrow made by the plough. Such a name (furrow) could be justified by the use of the constellation as an indicator of the farming activities at the end of summer and beginning of autumn in Mesopotamia some thousand years ago. We remark that it is not clear whether in ancient Mesopotamia this large constellation or part of it was really identified with a maiden, or at least with a goddess (may be Ishtar; ROGERS, 1998). The identification of the constellation with a maiden could be actually a Syrian adaptation of Shala (e.g. LAFFITTE, 2008), and one should conclude therefore that such an identification is not very old. In other words, the association of α Virginis with an ear is probably much older than that of the whole constellation with a maiden (Virgo), and we propose that it dates back to the beginning of Neolithic.

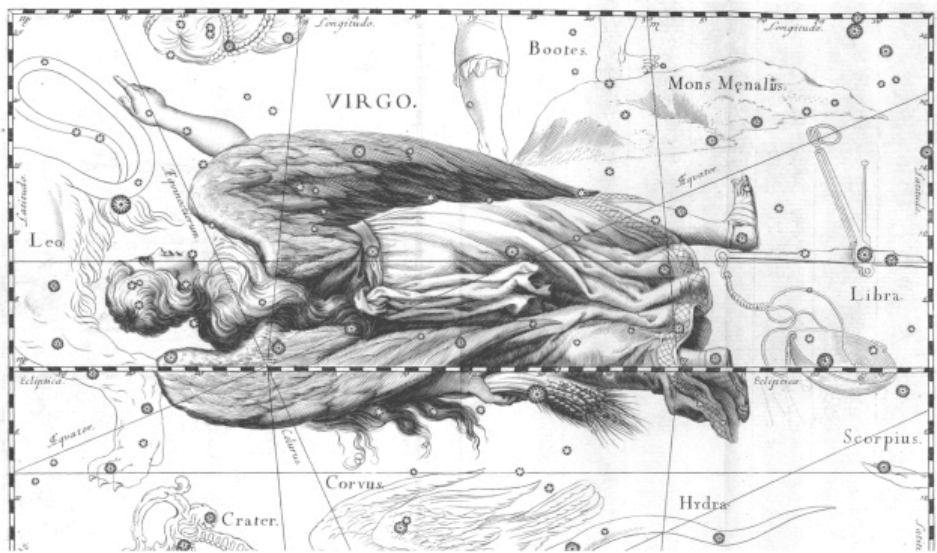

**Fig. 5.** The Virgo constellation in the Hevelius Atlas (INAF-Osservatorio Astronomico di Brera). Note the left hand with the ears, where Spica is located. This representation of the maiden is similar to that on the globe of the Farnese Atlas, the famous roman statue of the 2$^{nd}$ century CE.